\documentclass[twocolumn,prl,showpacs,multicol,amsmath,amssymb]{revtex4}
\usepackage[dvips]{graphicx}

\newcommand{\be}{\begin{equation}}
\newcommand{\ee}{\end{equation}}

\newcommand{\bea}{\begin{eqnarray}}
\newcommand{\eea}{\end{eqnarray}}
\newcommand{\bd}{\begin{displaymath}}
\newcommand{\ed}{\end{displaymath}}
\newcommand{\ba}{\begin{array}}
\newcommand{\ea}{\end{array}}
\newcommand{\bi}{\begin{itemize}}
\newcommand{\ei}{\end{itemize}}
\newcommand{\bc}{\begin{center}}
\newcommand{\ec}{\end{center}}
\newcommand{\bfl}{\begin{flushleft}}
\newcommand{\efl}{\end{flushleft}}
\newcommand{\bfr}{\begin{flushright}}
\newcommand{\efr}{\end{flushright}}

\def\6{\partial}

\def\no{\nonumber \\}

\def\={\!\!\!&=&\!\!\!}
\def\+{\!\!\!&&\!\!\!+~}
\def\-{\!\!\!&&\!\!\!-~}

\def\Y{$\rm{Yb  B_{12}}$}

\begin{document}
\date{15 January 2009}
\title{Theory of Spin  Exciton  in the Kondo Semiconductor $\rm{Yb  B_{12}}$ }
\author {Alireza Akbari$^{1,4}$, Peter Thalmeier$^{2}$, and  Peter Fulde$^{1,3}$ }
\affiliation{
$^{1}$Max Planck Institute for the Physcis of Complex Systems, D-01187 Dresden, Germany
\\$^2$Max Planck Institute for the  Chemical Physics of Solids, D-01187 Dresden, Germany
\\$^{3}$Asia Pacific Center for Theoretical Physics, Pohang, Korea
\\$^{4}$Institute for Advanced Studies in Basic Sciences,
45195-1159 Zanjan, Iran 
}
\begin{abstract}
The Kondo semiconductor \Y~exhibits  a spin and charge gap of  approximately 15 meV. Close to the gap energy narrow dispersive
collective excitations were identified by previous inelastic neutron scattering experiments. We present a 
theoretical analysis of these excitations. Starting from a periodic Anderson model for crystalline electric
field (CEF) split 4f states we derive the hybridized quasiparticle bands in slave boson mean-field approximation
and calculate the momentum dependent dynamical susceptibility in random phase approximation (RPA). We show that a small difference in the hybridization of 
the two CEF (quasi-) quartets leads to the appearance of two dispersive spin resonance excitations at the continuum threshold.
Their intensity is largest at the antiferromagnetic (AF) zone boundary point and they have an upward dispersion which merges with the continuum less than halfway into the Brillouin zone. Our theoretical analysis explains the most salient features of
previously unexplained experiments on the magnetic excitations of \Y.
\end{abstract}


\maketitle
The so-called Kondo insulators or semiconductors like, e.g., $\rm{CeNiSn}$, SmB$_6$  and $\rm{Yb  B_{12}}$
represent a special class of strongly correlated electrons\cite{Aeppli92}. 
In these compounds the 
 conduction electrons hybridize with  nearly localized 4f electrons.
The Coulomb
repulsion of the latter results in a  small energy gap\cite{Hundley90}  of order 10 meV  at the 
Fermi level\cite{Kasaya83,Iga98}.

At temperatures higher than the gap energy  these materials behave like
Kondo metals exhibiting their typical spin fluctuation spectrum. But a low temperatures a spin and charge gap opens
indicating the formation of an insulating singlet ground state\cite{Aeppli92,Riseborough00}.
This may be concluded from the total suppression of the local moment in
the susceptibility and from the semiconducting behavior of the resistivity, 
respectively \cite{Susaki99}. The gap formation may also be seen directly in the dynamical susceptibility
and finite frequency conductivity as probed in inelastic neutron scattering (INS)
and optical conductivity experiments. In cubic \Y~ the spin \cite{Nemkovski07} and charge \cite{Okamura05} 
gap obtained in this way are approximately equal to 15 meV but in general they need not be identical.

In addition unpolarized \cite{Mignot05} and polarized \cite{Nemkovski07} INS has found an 
interesting dispersive fine structure around this threshold energy.
Three excitation branches 
have been identified with energies $15$, $20$ and $38$ meV, respectively by analyzing  
the spectral function of the dynamical  susceptibility. Since the lower two INS peaks  are  narrow  and mostly 
centered  at  the  zone  boundary L-point with ${\bf  Q}=(\pi,\pi,\pi)$  they  may  be associated with the formation of  a  collective  heavy  quasiparticle spin resonance exciton appearing around 
the spin gap threshold~\cite{Mignot05,Nemkovski07} and driven by heavy quasiparticle interactions.
The collective modes remain visible in the $20$ meV region 
up to $T=159$ K \cite{Nefeodova99,Bouvet98}.
Similar spin resonance phenomena appear 
as result of    feedback effect in unconventional  heavy-fermion superconductors  below the
quasiparticle continuum threshold  at 2$\Delta_0$ where $\Delta_0$ is the gap amplitude\cite{Eremin08}.
The upper  peak is much broader and shows little dispersion. It is also rapidly suppressed 
with increasing temperature.  
It has been associated with  continuum excitations \cite{Mignot06} also
visible in a broad maximum in the optical conductivity\cite{Okamura05} around 38 meV.
   
Furthermore INS experiments on $\rm{Yb_{1-x}Lu_xB_{12}}$ compounds for different $\rm{Lu}$ concentrations
have  indicated that the disruption of coherence on the Yb
sublattice primarily affects the narrow peak structures occurring at $15-20$ meV
in pure $\rm{Yb  B_{12}}$  compound, whereas the spin gap and the broad magnetic signal around
38 meV remain almost unchanged\cite{Alekseev04}.

These intriguing experimental results have commonly been interpreted in a qualitative way within
the spin exciton scenario \cite{Mignot05,Nemkovski07,Riseborough01} but an alternative model was also
proposed \cite{Liu01}.
However no analysis of the former was attempted sofar although it is of fundamental importance to understand the microscopic origin and fine structure of the spin gap in Kondo semiconductors. In this communication we show in detail how the spin exciton bands in \Y~ arise on the background of a
single-particle continuum at the spin gap edge. We discuss the origin of the splitting into two modes, its
connection to CEF effects as well as their spectral shape and  dispersion. Our investigations clarify the underlying microscopic physics of these intriguing and for a long time unexplained observations.

Our starting point is the  hybridization-gap  
picture  based  on  the  periodic  Anderson model  which
is the most  widely  accepted    for the description of  Kondo  
semiconductors.  Using the mean-field slave  boson
approximation  for CEF split 4f states of Yb  we calculate the hybridized
bands.  With an empirical model for the quasiparticle interactions
we evaluate  the  momentum  
dependent  dynamical  magnetic  susceptibility in RPA. Its imaginary part is proportional to the INS
spectrum.  We  obtain sharp  resonance  features  around the continuum threshold and wave vectors
not too far from the zone  boundary L-point.  
Away  from  this  point  the  resonance peaks disperse upwards in energy and broaden. 
They merge  into  the  single  particle  continuum less than halfway into the Brillouin zone (BZ),  which  describes  the  basic  experimental  facts. In  addition  our  
calculation  suggests  that CEF splitting and associated CEF orbital dependence of hybridization 
are responsible for the observed splitting into two dispersive resonance modes. 

%
\begin{figure}
\centerline{\includegraphics[width=6cm,angle=0]{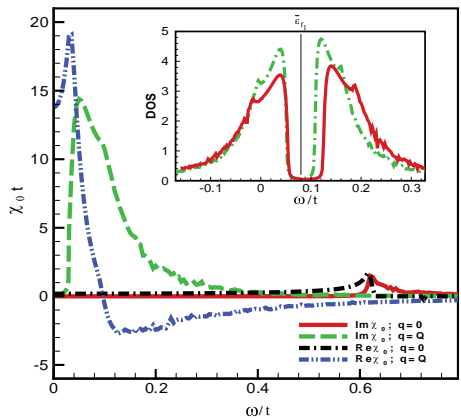}}
\caption{Dynamical susceptibility  in direction (1,1,1) for  
${\bf  q}=0$ (direct gap) and ${\bf  q}={\bf  Q}$ (indirect gap) versus  energy, 
for degenerate quasiparticle bands with $V_1=1t$ and  $\delta V=0$ ($\bar{V}=0.30t$, $J_{{\bf  Q}}=0$).
Inset shows the density of states for two CEF-split quasiparticle bands ($\bar{\epsilon}_{f_{1}}=0.08t$ and $b=0.41$).
The green curve corresponds to  the band  $V_{1}=t$ and the red one 
to the band $V_{2}=V_{1}+0.13t$. 
}
 \label{ImXqw_ND}
\end{figure}
%

The  $\rm{Yb}$ electronic configuration is $4f^{13}$ corresponding to a single hole
in the $4f$-shell\cite{Alekseev01}.
Therefore we consider the Anderson lattice model with a $f$-hole in a $j=7/2$ state, including the CEF effect, i.e.,
$
H_{t}=H_f+H_d+H_{f-d}+H_{C}.
$
Here $H_f$ describes the lattice of the localized, CEF-split $4f$-holes, $H_d$ the conduction electrons and 
$H_{f-d}$ is the hybridization  between both. Finally $H_{C}$ is 
the Coulomb interaction with an on-site hole repulsion $U_{ff}$. Our model assumes the limit $U_{ff}\rightarrow \infty$ where doubly occupied (hole) states (4f$^{12}$) are excluded and
the  two possible $\rm{Yb}$ configurations are either $4f^{14}$ or $4f^{13}$.
The one without   a $4f$ hole, i.e., $4f^{14}$ can be accounted
for by an auxiliary boson $b_i^{\dagger}$ \cite{Coleman84}. In cubic symmetry the $j = 7/2$ multiplet is split by the CEF
into  a quartet $\Gamma_8$ ground state and two excited doublet states.
 The latter may be treated as a quasi quartet $\Gamma_8'$ according to INS
results at higher temperatures\cite{Alekseev08}. The two quartets (index $\Gamma =1,2$) have energies $\Delta_{1}=0$ 
and $\Delta_{2} >0$.  The model Hamiltonian in the restricted zero- and one- hole Hilbert space is then
\begin{eqnarray}
 H &=&\sum_{i\gamma}(\epsilon_f+\Delta_{\gamma})f^\dagger_{i\gamma}f_{i\gamma}
+\sum_{{\bf  k}\gamma}\epsilon_{{\bf  k}}d^{\dagger}_{{\bf  k}\gamma}d_{{\bf  k}\gamma}
\no&+&
 N_s^{-1/2}\sum_{i{\bf  k}\gamma}(V_{{\bf  k}\gamma} e^{i{\bf  k }\cdot {\bf   R}_i} f^{\dagger}_{i\gamma}d_{{\bf  k}\gamma}b_i+c.c),
\label{model}
\end{eqnarray}
Here $\gamma=(\Gamma,m)$ where $\Gamma = 1,2$ denotes the quartet and $m=1-4$ is the orbital degeneracy index. Furthermore the local constraint 
$
\tilde{Q}_i=b_i^{\dagger}b_i+\sum_{\gamma}f^\dagger_{i\gamma}f_{i\gamma}=1
$
has to be respected for all $i$.
Therefore the total Hamiltonian including  the constraint is $H-\lambda_b\sum_{i}(\tilde{Q}_i-1)$, 
where $\lambda_b$ is  the Lagrange multiplier.
Here the $f^{\dagger}_{i\gamma}$ create $f$-holes at
lattice site $i$ in CEF state $\gamma$,
and the $d^{\dagger}_{{\bf  k}\gamma}$ create the  holes in the conduction band
with wave vector ${\bf  k}$ and CEF state index $\gamma$. 
The f-orbital energy is $\epsilon_f$, while $\Delta_{\gamma}=\Delta_{\Gamma}$ is the CEF excitation energy, and $N_s$ is the number of lattice sites. Finally $V_{{\bf  k}\gamma}$ is the hybridization energy between 4f  
and conduction holes. In the following, the ${\bf  k}$ dependence of the hybridization energy is neglected, i.e.,
$V_{{\bf  k}\gamma}=V_{\gamma}$. This is justified for a fully gapped Kondo insulator like \Y~ where $V_{{\bf  k}\gamma}$ does not vanish along lines in $\bf {k}$ space. Furthermore to use only a minimum set of model parameters, we replace $V_{\gamma} =V_{\Gamma,m}$ by $V_{\Gamma}=(1/2)(\sum_m|V_{\Gamma,m}|^2)^\frac{1}{2}$ which is the average over each set of quartet states.
We use a nearest-neighbor tight binding model with hopping $t$  for the  conduction electron bands $\epsilon_k$. The  spectral  function  of  the  experimental dynamical  susceptibility of  $\rm{Yb  B_{12}}$  exhibits two sharp peaks\cite{Nemkovski07}. Therefore it is essential that the two CEF quartets
have two different average hybridization energies $V_{\Gamma}$ ($\Gamma=1,2$).
%
\begin{figure}
\centerline{\includegraphics[width=7cm,angle=0]{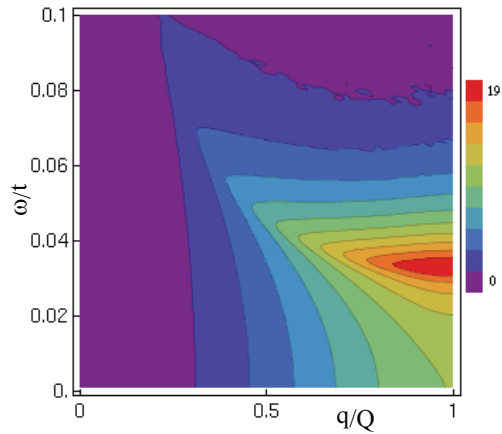}}
\caption{Contour plot of real part of noninteracting dynamical susceptibility for 
degenerate bands with  $V_1=1t$ and  $\delta V=0$ ($\bar{V}_1=0.30t$, $J_{{\bf  Q}}=0$) 
in the direction (1,1,1).
 }
\label{ReX0qw}
\end{figure}
%

The mean-field approximation to Eq.~(\ref{model}) is obtained by taking $b=\langle b_i\rangle$.
Minimizing the ground state energy with respect to 
 $b$  and the Lagrange multiplier $\lambda_b$ leads to the equations
\bea
\lambda_b b=\sum_{\gamma}V_{\gamma}W_{\gamma};\;\sum_{\gamma}n_{\gamma}^{f}+b^2=1;\;
n=\sum_{\gamma}(n_{\gamma}^{d}+n_{\gamma}^{f}),
\label{MF-eq}
\eea
where the following expectation values are introduced
$W_{\gamma}=\frac{1}{N_s}\sum_{{\bf   k}}\langle f^\dagger_{{ {\bf  k}}\gamma}d_{{ {\bf  k}}\gamma}\rangle$,
$n_{\gamma}^{d}=\frac{1}{N_s}\sum_{ {\bf  k}}\langle d^\dagger_{{ {\bf  k}}\gamma}d_{{ {\bf  k}}\gamma}\rangle$
and $n_{\gamma}^{f}=\frac{1}{N_s}\sum_{ {\bf  k}}\langle f^\dagger_{{ {\bf  k}}\gamma}f_{{ {\bf  k}}\gamma}\rangle$.
In Eq.~(\ref{MF-eq}), $n$ is the density of holes
 per site which
 defines the chemical potential $\mu$.
The mean-field Hamiltonian can be diagonalized. One obtains,
$
H_{MF}=
\sum_{{ {\bf  k}}\gamma,\pm} 
E_{\gamma,\pm}({\bf  k})a^\dagger_{{ {\bf  k}}\gamma,\pm}a_{{ {\bf  k}}\gamma,\pm},
$
where the hybridized bands have energies
$
E_{\gamma,\pm} ({\bf  k})=\frac{1}{2}\biggl[
\epsilon_{{\bf  k}}+\bar{\epsilon}_{f_{\gamma}}\pm\sqrt{(\epsilon_{{\bf  k}}-\bar{\epsilon}_{f_{\gamma}})^2+4\bar{V}^2_{\gamma}}\biggr]
$
which are still fourfold (m=1-4) degenerate.
The corresponding 4f-weight functions  of these quasiparticle bands are given by
$A_{\gamma,\pm}^{f}({\bf  k})=A_{\gamma,\mp}^{d}({\bf  k})=\frac{1}{2}[1\pm
\frac{\bar{\epsilon}_{f_{\gamma}}-\epsilon_{{\bf  k}}}{\sqrt{(\bar{\epsilon}_{f_{\gamma}}
-\epsilon_{{\bf  k}})^2+4\bar{V}_{\gamma}^2}}]$.
From the mean-field solution we also obtain
\bea
W_{\gamma}=
\frac{\bar{V}_{\gamma}}{N_s}\sum_{ {\bf  k}}
\frac{f(E_{\gamma,+} ({\bf  k}))-f(E_{\gamma,-} ({\bf  k}))}{E_{\gamma,+} ({\bf  k})-E_{\gamma,-} ({\bf  k})};
\no
n_{\gamma}^{f/d}=\frac{1}{N_s}\sum_{ {\bf  k},\pm}
A_{\gamma,\pm}^{f/d}(k)f(E_{\gamma,\pm} ({\bf  k})),
\eea
where $\bar{V}_{\gamma}=V_{\gamma}b$, and
$\bar{\epsilon}_{f_{\gamma}}=\epsilon_f+\Delta_{\gamma}-\lambda_b$.
In the zero temperature limit, $T=0$, the upper bands are empty.
Then the Fermi functions reduce to $f(E_{\gamma,+} ({\bf  k}))=0$,
and $f(E_{\gamma,-} ({\bf  k}))=\varphi_{\gamma}$ ($\sum_{\gamma}\varphi_{\gamma}=4n$).
Under the condition, $n=2$ or $\varphi_{\gamma}=1$, which holds as long as the chemical
potential is within the hybridization gap, 
we obtain the following mean-field equations from  Eqs.~(\ref{MF-eq}):
\bea
\bar{\epsilon}_{f_{1}}-\epsilon_f
=\sum_{\Gamma=1,2}
\frac{V_{\Gamma}^2}{2D}
\ln
\frac{
\sqrt{(D-\bar{\epsilon}_{f_{\Gamma}})^2+4\bar{V}_{\Gamma}^2}+D-\bar{\epsilon}_{f_{\Gamma}}
}{\sqrt{(D+\bar{\epsilon}_{f_{\Gamma}})^2+4\bar{V}_{\Gamma}^2}-D-\bar{\epsilon}_{f_{\Gamma}}};
\no\nonumber
2b^2=\sum_{\Gamma=1,2}
(\sqrt{(\bar{\epsilon}_{f_{\Gamma}}+D)^2+4\bar{V}_{\Gamma}^2}
-
\sqrt{(\bar{\epsilon}_{f_{\Gamma}}
-D)^2+4\bar{V}_{\Gamma}^2}).
\eea
%
Here $\bar{\epsilon}_{f_{2}}=\bar{\epsilon}_{f_{1}}+\Delta_{2}$, 
$\bar{V}_{2}=\bar{V}_{1}+\delta \bar{V}=b(V_{1}+\delta V)$, and $D=6t$ 
is half the conduction band width.
The density of states of the conduction band is assumed to be rectangular
($g(\epsilon)=1/2D;\;\;\;-D<\epsilon <D$ and zero otherwise). 
By solving the set of equations numerically one can find the $\bar{\epsilon}_{f_{1}}$ and $b$ values.
In order to be in the Kondo limit and have an insulating state with small hybridization gap the parameters should fulfill the condition $\Delta_{2}<\delta V<V_1,\mid \epsilon_f\mid <D$.
In the absence of CEF effects, by choosing $\epsilon_f=-0.75t$,  $V_{1}=t$,   $\delta V=0$, 
$\Delta_{2}=0$ we found $\bar{\epsilon}_{f_{1}}=0.05t$ and $b=0.30$
from the mean-field solutions which will be used in Fig.~\ref{ImXqw_ND}.

%
\begin{figure}
\centerline{\includegraphics[width=5cm,height=4cm,angle=0]{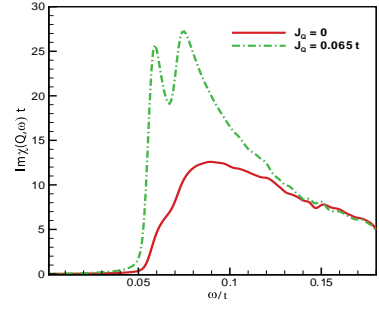}}
\caption{The  imaginary part of the susceptibility for the noninteracting ($J_{{\bf  Q}}=0$) 
and interacting case ($J_{{\bf  Q}_1}=J_{{\bf  Q}_2}=0.065t$) for $V_{1}=t$, ${\bf  q}={\bf  Q}$ and $\delta V=0.13 t$. 
The $J_{{\bf  Q}_\gamma}$  are slightly subcritical leading to a finite intrinsic resonance line width.}
 \label{ImX_Q}

\end{figure}
%

The dynamic magnetic susceptibility is calculated within RPA approximation.
Since we have two CEF quartets the spin response has the matrix form
$
\hat{\chi}({\bf  q},\omega)=[I-\hat{J}_{{\bf  q}}\hat{\chi}_0({\bf  q},\omega)]^{-1}\hat{\chi}_0({\bf  q},\omega),
$
where the unit matrix $I$,  the interaction $\hat{J}_{{\bf  q}}$ and the non-interacting quasiparticle susceptibility $\hat{\chi}_0({\bf  q},\omega)$ are $2\times 2$  matrices in the  CEF quartet index $\Gamma=1,2$.

The exchange interaction $\hat{J}_{{\bf  q}}$ between quasiparticles is assumed to be peaked at the 
AF wave vector ${\bf Q}=(\pi,\pi,\pi)$, i.e.,  the L-point because there the most pronounced
magnetic response is found. In principle  $\hat{J}_{{\bf  q}}$ may be calculated 
to order (1/$N^2$) \cite{Doniach87,Riseborough92}
but this is strongly model dependent. We choose to parameterize $\hat{J}_{{\bf  q}}$ 
in a simple way: The  interaction is peaked 
at {\bf Q} or  $ Q=\sqrt{3}\pi$  and it has the Lorentzian form
$J_{ {\bf  q}_{\Gamma\Gamma'}} =
\left[\frac{\Gamma_Q^2}{\left(
{\bf  q}-{\bf  Q}\right)^2+\Gamma_Q^2}\right]J_{ {\bf  Q}_{\Gamma\Gamma'}}$, where  $\Gamma_Q$ has the meaning of an inverse AF correlation length.
Each element  of the irreducible susceptibility matrix is calculated from the quasiparticle states as 
\cite{Riseborough92}:
\bea
\chi_0^{\Gamma\Gamma'}({\bf  q},\omega)
&=&\sum_{{\bf  k},\pm}A_{\Gamma,\pm}^{f}({\bf  k}+{\bf  q})A_{\Gamma',\mp}^{f}({\bf  k})\times
\no&&
\left[\frac{f(E_{\Gamma,\pm}({\bf  k}+{\bf  q}))-f(E_{\Gamma',\mp}({\bf  k}))}
{E_{\Gamma,\mp}({\bf  k})-E_{\Gamma',\pm}({\bf  k}+{\bf  q})-\omega}\right],
\eea
The non-diagonal elements of the interaction matrix corresponding to interactions of quasiparticles
with different CEF symmetry are neglected, implying 
$J_{ {\bf  q}_{\Gamma\Gamma'}}={\it J}_{\Gamma}({\bf  q})\delta_{\Gamma\Gamma'}$.
Then the RPA susceptibility is simply a sum of two contributions $\chi^{\Gamma\Gamma}({\bf  q},\omega)$ from the two sets of hybridized bands:
\be
\chi({\bf  q},\omega)=\sum_{\Gamma}[1-{\it J}_{\Gamma}({\bf  q})
\chi_0^{\Gamma\Gamma}({\bf  q},\omega)]^{-1}\chi_0^{\Gamma\Gamma}({\bf  q},\omega).
\ee

%
\begin{figure}
\centerline{\includegraphics[width=7cm,angle=0]{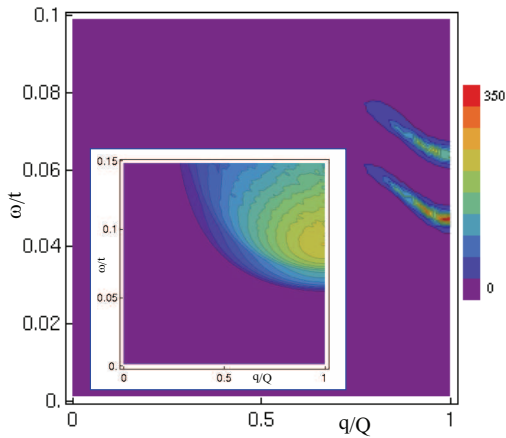}}
\caption{Contour plot of imaginary part of RPA dynamical susceptibility  with Lorentzian interaction 
$J_{\Gamma}(q)$ for CEF-split quasiparticle bands,  $\bar{\epsilon}_{f_{1}}=0.08t$ and $b=0.41$.
Here  $V_{1}=t$ and  $V_{2}=V_{1}+0.13t$
($J_{{\bf  Q}_1}=0.125t$, $J_{{\bf  Q}_2}=0.143t$ and $\Gamma_Q=2$ which satisfy the resonance condition in RPA formula: $J_{\Gamma}({\bf  Q})=1/Re[\chi^{\Gamma\Gamma}_0({\bf  Q},\omega_r)]$),
in direction (1,1,1).
The peaks  at the zone boundary (${\bf  q}={\bf  Q}$) appear at $\omega_1=0.047t$ and $\omega_2=0.063t$.
By choosing $t=0.32\rm{eV}$ (D = 1.92 eV) then $\omega_1=15\rm{meV}$ and $\omega_2=20\rm{meV}$ which are comparable with experimental results.
The inset shows the contour plot of imaginary part of dynamical susceptibility of 
noninteracting degenerate bands for comparison  ($V_1=1t$, $\delta V=0.13t$, $\bar{V}_1=0.41t$, and  $J_{{\bf  Q}}=0$) 
in the direction (1,1,1). The color scale of the inset is the same as Fig.\ref{ReX0qw}.}
\label{C-ImXqw_CEF}
\end{figure}
%

We now discuss the results of numerical calculations based on the previous analysis.
In  Fig.~(\ref{ImXqw_ND}) we have plotted the real and imaginary part of 
$\chi_0^{\Gamma\Gamma}({\bf  q},\omega)$ 
 without CEF splitting  ($\Delta_{2}=\delta V =0$) 
versus energy for wave vectors ${\bf  q}={\bf  0}$ and ${\bf  q}={\bf  Q}$.
One notices that $\rm{Im}\chi_0^{\Gamma\Gamma}({\bf  Q},\omega)$ has a strong low-energy peak due to
a small indirect gap while $\rm{Im}\chi_0^{\Gamma\gamma}({\bf  0},\omega)$
has a small peak at much higher energy due to a large direct gap. The broad structure of the former
 is due 
to noninteracting single-particle excitations
and the q and $\omega$ dependence is depicted in the inset of Fig.~\ref{C-ImXqw_CEF}.

The density of states for noninteracting quasiparticles
$\rho_{\gamma}(\omega)=\frac{1}{N_s}\sum_{{\bf  k},\pm}\delta(\omega-E_{\gamma,\pm}({\bf  k}))$ 
including the CEF splitting for the two sets of bands
with  $\Delta_2=0.01t$ and  $\delta V =0.13t$ is  plotted in the inset of Fig.~(\ref{ImXqw_ND}).
The two hybridization gaps are different due to a finite $\delta V$. However the latter is kept small
enough to ensure that the chemical potential is within the gap.

When the AF interaction ${\it J}_{\Gamma}({\bf  q})$  is turned on,
 the imaginary part becomes 
 $\rm{Im}\chi^{\Gamma\Gamma}({\bf  q},\omega)=F(\alpha_{\Gamma},\eta_{\Gamma})/{\it J}_{\Gamma}({\bf  q})$ where 
$\alpha_{\Gamma}={\it J}_{\Gamma}({\bf  q})\rm{Im}\chi^{\Gamma\Gamma}_0({\bf  q},\omega)$, $\eta_{\Gamma}=1-{\it J}_{\Gamma}({\bf  q}) Re\chi^{\Gamma\Gamma}_0({\bf  q},\omega)$, 
and $F(\alpha_{\Gamma},\eta_{\Gamma})=\alpha_{\Gamma}/(\eta_{\Gamma}^2+\alpha_{\Gamma}^2)$. 
In that case
the spectrum for q = Q moves to lower energies and a narrow  double-peak
structure, i.e., the collective spin resonance excitations appear.
Their energies $\omega_r^\Gamma$ are determined by the solution of $\eta_\Gamma =0$.
If they are lying within the indirect hybridization gap  one has $\alpha_\Gamma \rightarrow 0$.
Then the spectral function is a  delta function $\pi\delta(\eta_{\Gamma})/{\it J}_{\Gamma}({\bf  q})$
at the resonance energy $\omega_r$.
The dispersion  of the resonance, is determined by the real part of $\chi_0^{\Gamma\Gamma}({\bf  q},\omega)$ 
presented in Fig.~\ref{ReX0qw}. The plot shows that for $q<Q$  the 
maximum of the spectral function follows a ridge which decreases in height
and bends to higher energy. This turns into an upward dispersion of the resonance pole.
Its endpoint in the BZ is limited by the extension of the ridge in $Re\chi_0$.
The latter is fixed for the simple hybridization band model used here. A more realistic band model might give a larger extension
than the one seen in Fig.~\ref{ReX0qw}.

Due to the CEF effect the f-levels split into two (pseudo-) quartets ($\Delta_{2} > 0$) 
which hybridize {\em differently}. For $\delta V >0$ the resonance $\omega_r^{\Gamma=2}$
 associated with the $\gamma =2$ hybridized bands moves to higher energy and a second peak in addition to the 
 one at  $\omega_r^{\Gamma=1}$ appears in the spectral function. This is clearly seen in Fig.~\ref{ImX_Q} where
 the CEF split resonance peaks at ${\bf  q}={\bf  Q}$ appear around the threshold energy of the non-interacting continuum states. 
 In this figure we use subcritical values for the interaction constants.
Therefore the resonance peaks are right above the continuum
 threshold and have a finite intrinsic line width.
 
If the interaction constants are slightly increased the 
resonances move below the continuum and turn into true spin exciton poles without
intrinsic line widths (within RPA). Their dispersion is shown in the main panel of Fig.~\ref{C-ImXqw_CEF}. Away from the L-point ($Q=\sqrt{3}\pi$)  they disperse upwards and merge into the continuum. We identify these spin resonance modes with the observed experimental peaks at 15 and 20 meV \cite{Mignot05,Nemkovski07}  and we have chosen parameters such that their energy splitting and dispersion are reproduced. Our numerical  calculations show that the best fit to experiments  is obtained for $\delta V=0.13t$  where $\bar{\epsilon}_{f_{1}}=0.08t$ and $b=0.41$. 
These spin exciton peaks separate with increasing CEF splitting $\Delta_{2}$ and hybridization energy difference $\delta V$.
Therefore the influence of the latter 
is  strong since it directly affects the hybridization 
gap and hence the noninteracting susceptibility and resonance condition. 
We note that an increase in $J_{{\bf  Q}}$ (or a decrease of the hybridsation gap) will lead to a decrease of the spin exciton mode frequencies at {\bf Q}. For $J_{{\bf  Q}_1}=0.179 t$ the lowest mode would become soft. This softening signifies the instability of the paramagnetic state and the onset of AF order in a Kondo semiconductor. This is not observed in \Y~ at ambient pressure. We suggest that an investigation of the pressure dependence of spin exciton mode frequencies at {\bf Q} would give important clues how close \Y~ is to AF order. Finally we mention that our present model does not include the broad excitations at 38 meV. As has been suggested in Ref.~\onlinecite{Mignot06} it might be due to a continuum of additional band states which do not take place in the resonance formation. Their inclusion would require a multi-orbital  conduction band model.\\
We thank  P. A. Alekseev and I. Eremin for helpful discussions.


\end{document}